\documentclass[pra,aps,10pt,twocolumn,amsmath,amssymb,longbibliography]{revtex4-2}
\usepackage[english]{babel}
\usepackage{graphicx}
\usepackage{bm}
\usepackage{color}
\usepackage{slashed}
\usepackage{latexsym}
\usepackage{appendix}
\usepackage{dsfont}

\newcommand{\dd}{\mathrm{d}}
\newcommand{\ii}{\mathrm{i}} 
\newcommand{\ee}{\mathrm{e}} 

\newcommand{\me}{m_\mathrm{e}} 

\newcommand{\invf}{\raisebox{0.5em}{\rotatebox{180}{{\scriptsize F}}}}
\newcommand{\FeynmanState}[3]{#1_{\mathrm{F}{;{#2}}}^{(#3)}}
\newcommand{\AntiFeynmanState}[3]{#1_{\invf{;{#2}}}^{(#3)}}

\begin{document}

\title{Spirals, vortices, and helicity entanglements in dynamical Sauter-Schwinger pair creation}
\author{M. M. Majczak} 
\author{K. Krajewska} \email{Katarzyna.Krajewska@fuw.edu.pl}
\author{A. Bechler} 
\author{J. Z. Kami\'nski} \email{Jerzy.Kaminski@fuw.edu.pl}
\affiliation{
Institute of Theoretical Physics, Faculty of Physics, University of Warsaw, Pasteura 5, 02-093 Warsaw, Poland}
\date{\today}

\begin{abstract}
We study helicity correlations of electron-positron pairs created by a homogeneous time-dependent electric field in the Sauter-Schwinger scenario. 
Our analysis is based on solving the Dirac equation with the Feynman
or anti-Feynman boundary conditions, which is equivalent to the scattering matrix approach widely used in high energy physics. Most importantly, both these methods allow to fully 
account for the helicity (or, more generally, spin) correlations of created particles. The influence of helicity correlations and the carrier-envelope phase of the 
electric pulse on the properties of topological structures (such as spirals and vortices) in momentum distributions of created particles is investigated. The generation of maximally entangled helicity states is discussed 
and the possibility of using a short electric pulse as a fast switch between them is indicated.
\end{abstract}

\maketitle

\section{Introduction}
\label{sec:intro}

Modern relativistic quantum field theories are based on ideas and methods developed by Dirac~\cite{doi:10.1098/rspa.1928.0023}, 
Tomonaga~\cite{10.1143/PTP.1.27}, Feynman~\cite{feymann1949positrons,feymann1949qed}, Schwinger~\cite{schwinger1951gaugeinvariance}, and 
Dyson~\cite{dyson1949radiation,dyson1949s} (see, also, the original papers collected in the book~\cite{SchwingerQEDBook}). In particular, 
Schwinger reformulated achievements of Heisenberg and Euler~\cite{heisenberg1936diractheory}, and Sauter~\cite{sauter1931pairdirac} such that 
it became possible to account for the radiative corrections in quantum electrodynamics (QED). In consequence, this led to the complex effective 
Lagrangian, the imaginary part of which determines the probability density rates for the electron-positron pair creation by constant electromagnetic 
fields. Equally fundamental for the further development of the relativistic field theories in general and QED in particular, were the investigations 
of Feynman. They allowed to reinterpret the concept of anti-particles such that positrons are in fact electrons, but moving backward in time. 
These and related achievements can be found in the well-known Schweber's textbook~\cite{SchweberQEDBook}, or in the review 
articles~\cite{RevModPhys.58.449,ruffini2010pairastro,cajiao2019electron}.

The aim of this paper is to investigate the Sauter-Schwinger electron-positron pair creation by time-dependent electric field pulses. 
A frequently used method for analyzing this process is the approach based on the Dirac-Heisenberg-Wigner (DHW) function~\cite{bialynicki1991diracvacuum}
(see, also Ref.~\cite{bialynicki1991TheoryOfQuanta} for the nonrelativistic version). It is based on the second quantization formalism and leads to the 
so-called system of Quantum Kinetic Equations (QKE) for momentum and spin distributions of created 
electrons~\cite{hebenstreit2010schwinger,kohlfurst2020magneticpair,otto2015schwinger,PhysRevD.98.056009,PhysRevD.88.045017,dumlu2009pair,PhysRevA.100.012104,brodin2021plasmadynamics,blinne2014pair,blinne2016wigner}, 
but not for the correlated electron and positron spin distributions (see also the Computational Quantum Field 
Theory~\cite{doi:10.1080/00107510903450559} which is, however, beyond the scope of our presentation). It has been shown in 
Ref.~\cite{bechler2023schwinger} that, for spatially homogeneous electric fields, the DHW and QKE approaches follow directly from the Dirac equation 
provided that the proper boundary conditions are imposed on the solutions of this wave equation. Additionally, in order to get the same initial 
conditions (as only in this case both approaches lead to exactly same predictions), the solutions of the Dirac equation in the remote past for positrons 
have to be normalized to one. This extra requirement may raise some doubts, since in the Sauter-Schwinger process there are no real particles 
in the past. Both electrons and positrons appear in the form of virtual excitations of the quantum vacuum and, as such, do not have to be on the mass shell.
To solve this dilemma, in Ref.~\cite{PhysRevD.110.116025} an alternative approach to the dynamical Sauter-Schwinger process, based on the scattering 
matrix and reduction formulas, has been presented. As it results from the formalism developed there, the initial state is uniquely fixed by imposing 
strictly defined boundary conditions on the solutions of the Dirac equation. Namely, it must be the Feynman boundary condition when we 
determine the spin and momentum distributions of electrons in the far future for fixed spin and momentum of the created positron. Conversely, if we 
fix the spin and momentum of the created electron, then the spin-momentum distributions of the positrons are determined by such solutions of the 
Dirac equation that satisfy the anti-Feynman boundary conditions. Both these boundary conditions are going to be discussed in detail below. 
The initial condition defined in this way depends on the spins of the created particles and is not normalized to one. As it follows from the analysis 
presented in Ref.~\cite{PhysRevD.110.116025}, only for electric field strengths much smaller than the Schwinger critical value $\mathcal{E}_S$ 
this normalization is very close to one. Therefore, one concludes that for the electric field strength such that $|\bm{\mathcal{E}}(t)|\ll \mathcal{E}_S$ 
both approaches give nearly identical results for the momentum distributions summed over the spins of created particles. 
Note that the above boundary conditions, that strictly follow from the scattering matrix and reduction formulas, make difference between the relativistic 
quantum field theories and the condensed matter physics. For the latter, electrons do exist in the remote past in the form of the occupied valence bands 
(or the so-called Dirac sea), therefore their states can be normalized.

The approach based on the scattering matrix and reduction formulas provides the complete information about the complex probability amplitudes for 
the pair creation process and about the electron and positron spin correlations, in contrast to the DHW and QKE formalisms. This opens up new 
research perspectives for the strong-field QED in general (see, e.g., \cite{PhysRevD.109.076004,photonics12040307,Tang2025}) and for the creation 
of electron-positron pairs by classical electric fields in particular. For instance, problems related to the topological properties of helicity-momentum 
distributions and quantum entanglement of helicity states can be investigated now, as discussed below.

The plan of this paper is the following. In Sec.~\ref{sec:dirac} we show how the fully correlated electron-positron spin-resolved 
momentum distributions follow directly from the Dirac equation. This leads us to the exact same probability amplitudes and helicity-momentum distributions 
that can be derived by applying the more sophisticated approach based on the scattering matrix and reduction formulas, presented 
in Ref.~\cite{PhysRevD.110.116025}. The Feynman and anti-Feynman boundary conditions are introduced in Sec.~\ref{sec:feynman}, where we also 
demonstrate how the corresponding probability amplitudes can be evaluated by applying the transfer matrix discussed in Sec.~\ref{sec:transfer}. 
Topological structures in momentum distributions, such as spirals and vortex lines, are discussed in Sec.~\ref{sec:Spiralvortex}. We show there that 
although the existence of spiral distributions are not crucially influenced by the helicity correlations of the created pair, the occurrence of vortex 
lines significantly depends on these correlations and on the properties of the applied electric field. In particular, the change of the phase of 
the electric field strongly affects the shape of the vortex lines and, depending on the helicity degrees of freedom, can lead to their annihilation 
or flattening. Sec.~\ref{sec:Bellstates} is devoted to the analysis of quantum entanglement of helicity states of created pairs. We show that the 
generated entangled state can be effectively controlled by the applied electric field. In particular, a suitable change of electric pulses can be 
used as a fast switch between orthogonal entangled helicity states. Finally, our conclusions are presented in Sec.~\ref{sec:conclusions}.

In numerical analysis, we use the relativistic units in which $\hbar=c=\me=|e|=1$ and $e=-|e|$, where $\me$ and $e$ are the electron rest  mass and 
charge. The Schwinger value for the electric field strength, $\mathcal{E}_S$, is defined such that $\me c^2=|e|\mathcal{E}_S\lambdabar_C$, 
where $\lambdabar_C=\hbar/\me c$ is the reduced Compton wavelength. Additionally, the Compton time is defined as $t_C=\hbar/\me c^2$. 
For the $\gamma$ matrices we apply the Dirac representation, use the Feynman's notation $\slashed{a}=a_\mu\gamma^\mu$, and the metric $(+,-,-,-)$. 
In analytical formulas we put $\hbar=1$ but keep explicitly $\me$, $e$, and $c$.

\section{Solutions of the Dirac equation}
\label{sec:dirac}

We start by introducing the basic definitions and notation that will be used throughout the paper. Specifically, we define normalized free bispinors,
\begin{equation}
 u^{(+)}_{\bm{p},\lambda}=\sqrt{\frac{p^{0}+\me c}{2p^{0}}}
\left(
\begin{array}{c}
 \chi_{_\lambda}  \\ 
 \frac{ \bm{\sigma}\cdot \bm{p}}{p^{0}+\me c}\chi_{_\lambda}
 \end{array}
\right),
\label{ewa1}
\end{equation}
\begin{equation}
 u^{(-)}_{-\bm{p},\lambda}=\sqrt{\frac{p^{0}+\me c}{2p^{0}}}
\left(
\begin{array}{c}
 \frac{ -\bm{\sigma}\cdot \bm{p}}{p^{0}+\me c}\chi_{_\lambda}\\
\chi_{_\lambda}
 \end{array}
\right),
\label{ewa2}
\end{equation}
where $p^{0}=\sqrt{ \bm{p}^{2}+(\me c)^{2}}$, $\lambda=\pm$, and
\begin{equation}
\chi_{+}=
\left(
\begin{array}{c}
\cos\frac{\theta_{s}}{2}\ee^{-\ii\varphi_{s}/2}\\
\sin\frac{\theta_{s}}{2}\ee^{\ii\varphi_{s}/2}
\end{array}\right), 
\,
\chi_{-}=
\left(
\begin{array}{c}
-\sin\frac{\theta_{s}}{2}\ee^{-\ii\varphi_{s}/2}\\
\cos\frac{\theta_{s}}{2}\ee^{\ii\varphi_{s}/2}
\end{array}
\right).
\label{ewa3}
\end{equation}
If $\bm{n}=\left(\sin{\theta_{s}}\cos{\varphi_{s}}, \sin{\theta_{s}}\sin{\varphi_{s}}, \cos{\theta_{s}}\right)$ is the axis of spin quantization, 
then we have
\begin{equation}
 (\bm{\sigma}\cdot \bm{n})\chi_{\pm}=\pm\chi_{\pm}.
\label{ewa4}
\end{equation} 
In particular, if $\theta_s$ and $\varphi_s$ are the polar and azimuthal angles of $\bm{p}$, respectively (or, in other words, if 
$\bm{p}=|\bm{p}|\bm{n}$), we recover the helicity states for electrons [Eq.~\eqref{ewa1}] and positrons [Eq.~\eqref{ewa2}].
Such normalized bispinors are solutions of the system of algebraic equations,
\begin{equation}
\left(\slashed{p}-\beta \me c\right) u^{(\beta)}_{ \bm{p},\lambda}=0,\quad \beta=\pm \, .
\label{ewa5}
\end{equation}
The four bispinors [Eqs.~\eqref{ewa1} and~\eqref{ewa2}] form the orthonormal basis, as they fulfill the normalization and completeness conditions,
\begin{equation}
[u^{\left(\beta\right)}_{\beta \bm{p},\lambda}]^{\dagger} u^{\left(\beta'\right)}_{\beta' \bm{p},\lambda'}=\delta_{\beta\beta'}\delta_{\lambda\lambda'}
\label{ewa6}
\end{equation}
\begin{equation}
\sum_{\beta,\lambda=\pm} u^{\left(\beta\right)}_{\beta\bm{p},\lambda}[u^{\left(\beta\right)}_{\beta \bm{p},\lambda}]^{\dagger}=\mathbb{I},
\label{ewa7}
\end{equation}
where $\mathbb{I}$ is the $4\times 4$ identity matrix. Moreover, the bispinors with $\beta=+$ describe free electrons, whereas those for $\beta=-$ 
correspond to positrons.

For our further purposes, we order these bispinors as follows,
\begin{equation}
 w_{ \bm{p},1}= u^{(+)}_{ \bm{p},+},\, w_{ \bm{p},2}= u^{(+)}_{ \bm{p},-},\, w_{ \bm{p},3}= u^{(-)}_{- \bm{p},+},\, w_{ \bm{p},4}= u^{(-)}_{- \bm{p},-}.
\label{ewa8}
\end{equation}
This means that we establish the equivalence,
\begin{equation}
w_{\bm{p},j}=u^{(\beta)}_{\beta\bm{p},\lambda},\quad j=1,\dots, 4,\quad (\beta,\lambda)=(\pm,\pm),
\label{ewa8a}
\end{equation}
with the unique relation $j \leftrightarrow (\beta,\lambda)$. Next, we introduce the unitary matrix, 
\begin{equation}
B_{ \bm{p}}=\left( w_{ \bm{p},1},\: w_{ \bm{p},2},\: w_{ \bm{p},3},\: w_{ \bm{p},4}\right),
\label{ewa9}
\end{equation} 
which relates the bispinor basis $w_{\bm{p},j}$ with the canonical one $e_{j}$ ($j=1,\:2,\:3,\:4)$,
\begin{equation}
 w_{ \bm{p},j}=B_{ \bm{p}}\cdot e_{j},\quad B^{\dagger}_{ \bm{p}}B_{ \bm{p}}=\mathbb{I}.
\label{ewa10}
\end{equation} 
The elements of the latter are columns with zeros everywhere except for the $j$-th row, whose value is 1. This basis transformation will turn out 
particularly useful when the time-evolution matrix for the Dirac equation will be constructed.

\subsection{Dirac equation in a time-dependent electric field}
\label{sec:DE}

Our aim is to analyze solutions of the Dirac equation in the time-dependent electric field $ \bm{\mathcal{E}}\left(t\right)$,
\begin{equation}
\ii \partial_{t}\psi\left( \bm{x},t\right)=\gamma^0\left[c \bm{\gamma}\cdot\left(-\ii\bm{\nabla}-e \bm{A}(t)\right)+\me c^2\right]\psi( \bm{x},t),
\label{ewa11}
\end{equation}
where the corresponding vector potential is defined as
\begin{equation}
 \bm{A}(t)=\int_{t}^{\infty}d\tau  \bm{\mathcal{E}}(\tau).
\label{ewa12}
\end{equation}
In this case, we seek solutions of Eq.~\eqref{ewa11} in the form of plane waves,
\begin{equation}
\psi( \bm{x},t)=\ee^{\ii  \bm{p}\cdot \bm{x}}\psi_{ \bm{p}}(t).
\label{ewa13}
\end{equation}  
This leads to the system of four complex ordinary differential equations,
\begin{equation}
\ii \dot{\psi}_{ \bm{p}}(t)=H_D(\bm{p},t)\psi_{ \bm{p}}(t),
\label{ewa14}
\end{equation}
where the dot over $\psi$ means the time-derivative and the time-dependent Hermitian matrix $H_D(\bm{p},t)$ is equal to
\begin{equation}
H_D(\bm{p},t)=\gamma^0\left[c \bm{\gamma}\cdot\left(\bm{p}-e \bm{A}(t)\right)+\me c^2\right].
\label{ewa15}
\end{equation} 
In general, we assume that 
\begin{equation}
\lim_{t \rightarrow \pm \infty}  e\bm{A}(t)=\bm{p}_{\pm}.
\label{ewa16}
\end{equation}
In practice, however, the above limits mean that there are times $t_\mathrm{i} < t_\mathrm{f}$ such that $e\bm{A}(t)=\bm{p}_-$ 
for $t < t_\mathrm{i}$ and $e\bm{A}(t)=\bm{p}_+$ for $t > t_\mathrm{f}$, which is equivalent to saying that for $t < t_\mathrm{i}$ 
and $t > t_\mathrm{f}$ the electric field vanishes, $\bm{\mathcal{E}}(t)=0$. Hence, we can define the asymptotic free-states,
\begin{equation}
\psi^{(\beta)}_{\bm{p},\lambda}(\bm{x},t)=\exp[-\ii\beta E_{\bm{p}-\bm{p}_-}(t-t_{\mathrm{i}})+\ii\bm{p}\cdot\bm{x}]u^{(\beta)}_{\beta(\bm{p}-\bm{p}_-),\lambda},
\label{ewa16a}
\end{equation}
for $t\leqslant t_{\mathrm{i}}$ and
\begin{equation}
\psi^{(\beta)}_{\bm{p},\lambda}(\bm{x},t)=\exp[-\ii\beta E_{\bm{p}-\bm{p}_+}(t-t_{\mathrm{f}})+\ii\bm{p}\cdot\bm{x}]u^{(\beta)}_{\beta(\bm{p}-\bm{p}_+),\lambda},
\label{ewa16b}
\end{equation}
for $t\geqslant t_{\mathrm{f}}$, where $E_{\bm{p}}=c\sqrt{\bm{p}^2+(\me c)^2}$. For $\beta=+$ they correspond to electrons of canonical 
momentum $\bm{p}$ and polarizations $\lambda=\pm$, for $\beta=-$ they describe positrons of canonical momentum $-\bm{p}$ and polarizations 
$\lambda=\pm$, whereas $\bm{p}-\bm{p}_{\pm}$ relate to the kinetic momenta. In the theoretical discussion presented below the name 'momentum' 
always refers to the canonical momentum. In the numerical analysis, however, we will choose electric field pulses such that $\bm{p}_{\pm}=\bm{0}$. 
Hence, the canonical and kinetic momenta are asymptotically identical.

\subsection{Evolution matrix}
\label{sec:evolution}

In order to find a general solution of Eq.~\eqref{ewa14} we proceed in the standard way. Namely, we introduce the unitary evolution matrix $U_D(\bm{p},t,t')$ that fulfills the differential equation,
\begin{equation}
\ii\frac{\dd}{\dd t}U_D(\bm{p},t,t')=H_D(\bm{p},t)U_D(\bm{p},t,t'),
\label{ewa17}
\end{equation}
with the initial condition $U_D(\bm{p},t',t')=\mathbb{I}$. Knowing the evolution matrix, one can propagate in time any initial state $\psi_{\bm{p}}(t_{\mathrm{i}})$,
\begin{equation}
\psi_{\bm{p}}(t)=U_D(\bm{p},t,t_{\mathrm{i}})\psi_{\bm{p}}(t_{\mathrm{i}}),
\label{ewa18}
\end{equation}
with the conservation of normalization, $[\psi_{\bm{p}}(t)]^\dagger\psi_{\bm{p}}(t)=[\psi_{\bm{p}}(t_{\mathrm{i}})]^\dagger\psi_{\bm{p}}(t_{\mathrm{i}})$.

Next, in order to get a more clear physical interpretation of elements of the evolution matrix $U_D(\bm{p},t_{\mathrm{f}},t_{\mathrm{i}})$,
we introduce another matrix,
\begin{equation}
U_{\bm{p}}(t_{\mathrm{f}},t_{\mathrm{i}})=B_{\bm{p}-\bm{p}_+}^\dagger U_D(\bm{p},t_{\mathrm{f}},t_{\mathrm{i}})B_{\bm{p}-\bm{p}_-}.
\label{ewa19}
\end{equation}
It is also unitary, as it follows from the properties~\eqref{ewa10}. However, only if $\bm{p}_-=\bm{p}_+$ 
it satisfies the initial condition $U_{\bm{p}}(t_{\mathrm{i}},t_{\mathrm{i}})=\mathbb{I}$. Note that this matrix explicitly depends on asymptotic 
momenta $\bm{p}_{\pm}$, but in order to simplify the notation we shall disregard this dependence in our further analysis.
According to the relations~\eqref{ewa10} between the canonical basis and solutions of the free Dirac equation, the elements of the matrix 
$U_{\bm{p}}$ in the canonical basis have the form,
\begin{equation}
	[U_{\bm{p}}(t_{\mathrm{f}},t_{\mathrm{i}})]_{jk}=w^\dag_{\bm{p}-\bm{p}_+,j}U_D(\bm{p},t_{\mathrm{f}},t_{\mathrm{i}})w_{\bm{p}-\bm{p}_-,k}.
\label{ewa19a}
\end{equation}
This shows that columns of $U_{\bm{p}}$ have clear physical interpretation. For instance, the first column with elements 
$[U_{\bm{p}}(t_{\mathrm{f}},t_{\mathrm{i}})]_{j1}$, corresponding to the initial electron state $u_{\bm{p}-\bm{p}_-,+}^{(+)}$ [cf. ordering (9)], 
contains probability amplitudes at time $t_{\mathrm{f}}$ for electrons of momentum $\bm{p}$ and spin polarizations $\pm$ (first and second rows), 
and for positrons of momentum $-\bm{p}$ and spin polarizations $\pm$ (third and fourth rows). In a similar way the second column contains probability amplitude for the initial electron state $u_{\bm{p}-\bm{p}_-,-}^{(+)}$, whereas third and fourth columns correspond to positron initial states $u_{-(\bm{p}-\bm{p}_-),+}^{(-)}$ and $u_{-(\bm{p}-\bm{p}_-),-}^{(-)}$, respectively. 
In general, if the initial state $\psi_{\bm{p}}(t_{\mathrm{i}})$ has the form
\begin{equation}
\psi_{\bm{p}}(t_{\mathrm{i}})=\sum_{\beta,\lambda=\pm}c^{(\beta)}_{\bm{p},\lambda}(t_{\mathrm{i}})u^{(\beta)}_{\beta (\bm{p}-\bm{p}_-),\lambda},
\label{ewa20}
\end{equation}
then the final state $\psi_{\bm{p}}(t_{\mathrm{f}})$ is equal to
\begin{equation}
\psi_{\bm{p}}(t_{\mathrm{f}})=\sum_{\beta,\lambda=\pm}c^{(\beta)}_{\bm{p},\lambda}(t_{\mathrm{f}})u^{(\beta)}_{\beta (\bm{p}-\bm{p}_+),\lambda}.
\label{ewa21}
\end{equation}
The corresponding amplitudes are related by
\begin{equation}
\mathbb{C}_{\bm{p}}(t_{\mathrm{f}})=U_{\bm{p}}(t_{\mathrm{f}},t_{\mathrm{i}})\mathbb{C}_{\bm{p}}(t_{\mathrm{i}}),
\label{ewa22}
\end{equation}
where we have introduced the notation,
\begin{equation}
\mathbb{C}_{\bm{p}}(t)=\begin{pmatrix} \mathbb{C}_{\bm{p}}^{(+)}(t) \cr \mathbb{C}_{\bm{p}}^{(-)}(t)\end{pmatrix}, \quad
\mathbb{C}_{\bm{p}}^{(\beta)}(t)=\begin{pmatrix} c^{(\beta)}_{\bm{p},+(}t) \cr c^{(\beta)}_{\bm{p},-}(t) \end{pmatrix},
\label{ewa23}
\end{equation}
with complex amplitudes $c^{(\beta)}_{\bm{p},\lambda}(t)$ such that
\begin{equation}
\sum_{\beta,\lambda=\pm}|c^{(\beta)}_{\bm{p},\lambda}(t)|^2=1,
\label{ewa24}
\end{equation}
for all times, as the evolution matrix is unitary.

A complementary meaning of the elements of $U_{\bm{p}}(t_{\mathrm{f}},t_{\mathrm{i}})$ can be gained from the time-evolution of the system. 
First, we define the propagator $U(\bm{x},t;\bm{x}',t')$ which fulfills the equation,
\begin{equation}
[\ii\partial_t-H_D(-\ii\bm{\nabla},t)]U(\bm{x},t;\bm{x}',t')=0,\quad t \geqslant t',
\label{ewa24a}
\end{equation}  
with the initial condition,
\begin{equation}
U(\bm{x},t';\bm{x}',t')=\delta^{(3)}(\bm{x}-\bm{x}').
\label{ewa24b}
\end{equation}  
Then, the space-time evolution of any initial state $\psi_{\mathrm{i}}(\bm{x},t)$ is given by the space-integral,
\begin{equation}
\psi(\bm{x},t)=\int\dd^3x'\, U(\bm{x},t;\bm{x}',t')\psi_{\mathrm{i}}(\bm{x}',t').
\label{ewa24c}
\end{equation}  
Hence, the transition amplitude to the final state $\psi_{\mathrm{f}}(\bm{x},t)$ within the time-interval $[t,t']$ is equal to
\begin{equation}
\mathcal{A}_{\mathrm{fi}}(t,t')=\int\dd^3x\,\dd^3x'\, [\psi_{\mathrm{f}}(\bm{x},t)]^{\dagger}U(\bm{x},t;\bm{x}',t')\psi_{\mathrm{i}}(\bm{x}',t').
\label{ewa24d}
\end{equation}  
Next, since the interaction depends only on time, therefore
\begin{align}
U(\bm{x},t_{\mathrm{f}};&\bm{x}',t_{\mathrm{i}})=\int\frac{\dd^3p}{(2\pi)^3}\ee^{\ii\bm{p}\cdot\bm{x}}U_D(\bm{p},t_{\mathrm{f}},t_{\mathrm{i}})\ee^{-\ii\bm{p}\cdot\bm{x}'} \\
=&\int\frac{\dd^3p}{(2\pi)^3}\ee^{\ii\bm{p}\cdot\bm{x}}B_{\bm{p}-\bm{p}_+}U_{\bm{p}}(t_{\mathrm{f}},t_{\mathrm{i}})[B_{\bm{p}-\bm{p}_-}]^{\dagger}\ee^{-\ii\bm{p}\cdot\bm{x}'}, \nonumber
\label{ewa24e}
\end{align}  
and the transition amplitude between the initial stationary state 
$\psi^{(\beta_{\mathrm{i}})}_{\bm{p}_{\mathrm{i}},\lambda_{\mathrm{i}}}(\bm{x}',t_{\mathrm{i}})$ [see, Eq.~\eqref{ewa16a}] and the final one 
$\psi^{(\beta_{\mathrm{f}})}_{\bm{p}_{\mathrm{f}},\lambda_{\mathrm{f}}}(\bm{x}',t_{\mathrm{f}})$ [see, Eq.~\eqref{ewa16b}] equals
\begin{equation}
\mathcal{A}_{\mathrm{fi}}=(2\pi)^3\delta^{(3)}(\bm{p}_{\mathrm{f}}-\bm{p}_{\mathrm{i}})[U_{\bm{p}_{\mathrm{i}}}(t_{\mathrm{f}},t_{\mathrm{i}})]_{j_{\mathrm{f}}j_{\mathrm{i}}},
\label{ewa24f}
\end{equation}  
where $j_{\mathrm{i}}\leftrightarrow (\beta_{\mathrm{i}},\lambda_{\mathrm{i}})$ and 
$j_{\mathrm{f}}\leftrightarrow (\beta_{\mathrm{f}},\lambda_{\mathrm{f}})$, according to the ordering~\eqref{ewa8a}. 
Note that momenta for particles and anti-particles have opposite space directions.

\begin{figure}
\includegraphics[width=7.5cm]{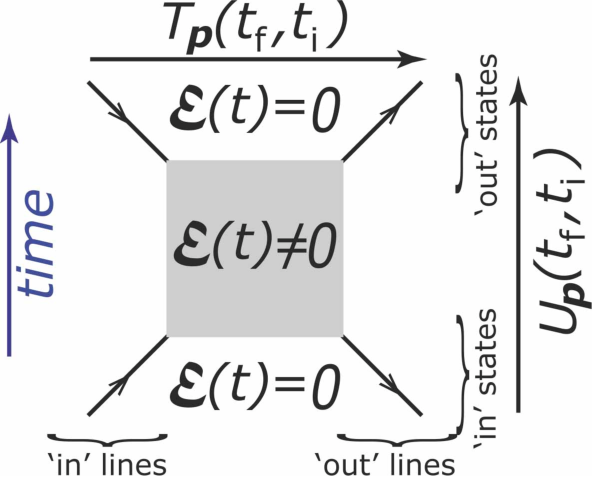}
\caption{Graphical representation of the time-evolution vs. `line-evolution' for the Feynman diagram with four external fermion lines.
}
\label{PairFeynman}
\end{figure}

In the definition~\eqref{ewa19} we have selected times such that either before $t_{\mathrm{i}}$ or after $t_{\mathrm{f}}$ the electric field vanishes. 
Our choice is motivated by future applications of the theory developed here in which the time-dependent electric field exists only for a finite time 
and the pair creation is assisted by the entire electric field pulse. If, however, one is interested in the investigation of the process 
within the electric field pulse, then the more general form of the evolution matrix $U_{\bm{p}}(t_{\mathrm{f}},t_{\mathrm{i}})$ could be used,
\begin{equation}
U_{\bm{p}}(t,t')=B_{\bm{p}-e\bm{A}(t)}^\dagger U_D(\bm{p},t,t')B_{\bm{p}-e\bm{A}(t')},
\label{ewa25}
\end{equation}
in which temporal basis of the Dirac bispinors, Eqs.~\eqref{ewa1} and~\eqref{ewa2}, are applied.

\subsection{Transfer matrix}
\label{sec:transfer}

Since the Feynman's pioneering works~\cite{feymann1949qed,feymann1949positrons}, physical processes of relativistic quantum field theories are predominantly 
described in terms of the Feynman diagrams. In QED these diagrams are build off fermion and photon lines, and the external lines that go to infinities 
represent the real particles. If we choose the time arrow vertically in the 'up-direction', then the fermion lines can be oriented either upwards or 
downwards: the upward direction describes electrons and the downward one corresponds to positrons. Since for the pair creation process in the final 
state we have both the electron and positron lines, and in addition we do not have external photon lines, as the time-dependent electric field is 
treated classically, therefore, also in the initial state we have two lines, one for electrons and another one for positrons. Hence, the initial 
state for the time-evolution consists of incoming (electron) and outgoing (positron) lines. Similar situation is met for the final state, i.e., outgoing 
line represents the particle (electron) and the incoming one corresponds to the anti-particle (positron). However, one can look at the Feynman 
diagram in the horizontal direction and try to relate the incoming lines to the outgoing ones. While in the vertical direction such a process 
is described by the evolution matrix, in the horizontal direction the transformation of the incoming to the outgoing lines is given by another matrix, 
which we call below the transfer matrix.

The transfer matrix $T_{\bm{p}}(t,t')$ relates the incoming lines to the outgoing ones, i.e.,
\begin{equation}
\begin{pmatrix} \mathbb{C}_{\bm{p}}^{(+)}(t) \cr \mathbb{C}_{\bm{p}}^{(-)}(t')\end{pmatrix}
=T_{\bm{p}}(t,t')\begin{pmatrix} \mathbb{C}_{\bm{p}}^{(+)}(t') \cr \mathbb{C}_{\bm{p}}^{(-)}(t)\end{pmatrix}.
\label{ewa26}
\end{equation}
It can be derived from the evolution matrix $U_{\bm{p}}(t,t')$. To this end let us write
\begin{equation}
U_{\bm{p}}(t,t')=
\begin{pmatrix} U_{\bm{p}}^{(++)}(t,t') & U_{\bm{p}}^{(+-)}(t,t') \cr
U_{\bm{p}}^{(-+)}(t,t') & U_{\bm{p}}^{(--)}(t,t') \end{pmatrix}
\label{ewa27}
\end{equation}
and
\begin{equation}
T_{\bm{p}}(t,t')=
\begin{pmatrix} T_{\bm{p}}^{(++)}(t,t') & T_{\bm{p}}^{(+-)}(t,t') \cr
T_{\bm{p}}^{(-+)}(t,t') & T_{\bm{p}}^{(--)}(t,t') \end{pmatrix},
\label{ewa28}
\end{equation}
where $U_{\bm{p}}^{(\beta\beta')}(t,t')$ and $T_{\bm{p}}^{(\beta\beta')}(t,t')$ ($\beta,\beta'=\pm$) are $2\times 2$ matrices. Then, after some algebra, we find,
\begin{align}
T_{\bm{p}}^{(++)}(t,t')=&U_{\bm{p}}^{(++)}(t,t') \nonumber \\
                       &-U_{\bm{p}}^{(+-)}(t,t')[U_{\bm{p}}^{(--)}(t,t')]^{-1}U_{\bm{p}}^{(-+)}(t,t'),
\nonumber \\
T_{\bm{p}}^{(+-)}(t,t')=&U_{\bm{p}}^{(+-)}(t,t')[U_{\bm{p}}^{(--)}(t,t')]^{-1},
\nonumber \\
T_{\bm{p}}^{(-+)}(t,t')=&-[U_{\bm{p}}^{(--)}(t,t')]^{-1}U_{\bm{p}}^{(-+)}(t,t'),
\nonumber \\
T_{\bm{p}}^{(--)}(t,t')=&[U_{\bm{p}}^{(--)}(t,t')]^{-1}.
\label{ewa29}
\end{align}
Similarly, the inverse of the transfer matrix relates the outgoing lines to the incoming ones,
\begin{equation}
\begin{pmatrix} \mathbb{C}_{\bm{p}}^{(+)}(t') \cr \mathbb{C}_{\bm{p}}^{(-)}(t)\end{pmatrix}
=[T_{\bm{p}}(t,t')]^{-1}\begin{pmatrix} \mathbb{C}_{\bm{p}}^{(+)}(t) \cr \mathbb{C}_{\bm{p}}^{(-)}(t')\end{pmatrix}.
\label{ewa30}
\end{equation}
Although the formulas derived above are valid for arbitrary times $t>t'$, we will apply them for $t=t_{\mathrm{f}}$ and $t'=t_{\mathrm{i}}$. 
A schematic representation of the time-evolution (given by the evolution matrix) and the `line-evolution' (given by the transfer matrix) 
is shown in Fig.~\ref{PairFeynman}.
Note that a similar method, but for the transfer and the scattering matrices, has been used 
in investigations of electron transport phenomena in nanostructures~\cite{KAMINSKI2011251,https://doi.org/10.1002/andp.201200162}.

\section{Feynman and anti-Feynman boundary conditions}
\label{sec:feynman}

Investigation of dynamics in nonrelativistic quantum mechanics consists in finding solutions of the Schr\"odinger equation that satisfy the retarded 
or advanced boundary conditions. However, in relativistic quantum theories, because we deal with free-particle states of negative energy, there exists 
two extra boundary conditions, which are called the Feynman and anti-Feynman ones~\cite{bialynicki1975quantumelectro}. Let us stick further to 
the physical problem considered in this paper, i.e., to the evolution of solutions of the Dirac equation in the time-dependent electric field that 
vanishes in the remote past and far future, as described above. We say that the solution $\FeynmanState{\psi}{\bm{p},\lambda_0}{-}(\bm{x},t)$ fulfills 
the Feynman boundary condition if
\begin{equation}
\FeynmanState{\psi}{\bm{p},\lambda_0}{-}(\bm{x},t)=
\begin{cases}
\phi^{(-)}_{\bm{p},\lambda_0}(\bm{x},t)+\sum_{\lambda=\pm}\phi^{(+)}_{\bm{p},\lambda}(\bm{x},t),
& t>t_{\mathrm{f}}, \cr 
\sum_{\lambda=\pm}\phi^{(-)}_{\bm{p},\lambda}(\bm{x},t), & t<t_{\mathrm{i}},
\end{cases}
\label{ewa31}
\end{equation}
where [see, Eqs.~\eqref{ewa16a} and \eqref{ewa16b}]
\begin{align}
\phi^{(-)}_{\bm{p},\lambda_0}(\bm{x},t)&=\ee^{\ii E_{\bm{p}-\bm{p}_+}(t-t_{\mathrm{f}})}\ee^{\ii\bm{p}\cdot\bm{x}}u^{(-)}_{-(\bm{p}-\bm{p}_+),\lambda_0} ,
\nonumber \\
\phi^{(+)}_{\bm{p},\lambda}(\bm{x},t)&=\ee^{-\ii E_{\bm{p}-\bm{p}_+}(t-t_{\mathrm{f}})}\ee^{\ii\bm{p}\cdot\bm{x}}c^{(+)}_{\bm{p},\lambda}(t_{\mathrm{f}})u^{(+)}_{\bm{p}-\bm{p}_+,\lambda} ,
\nonumber \\
\phi^{(-)}_{\bm{p},\lambda}(\bm{x},t)&=\ee^{\ii E_{\bm{p}-\bm{p}_-}(t-t_{\mathrm{i}})}\ee^{\ii\bm{p}\cdot\bm{x}}c^{(-)}_{\bm{p},\lambda}(t_{\mathrm{i}})u^{(-)}_{-(\bm{p}-\bm{p}_-),\lambda},
\label{ewa32}
\end{align}
and $E_{\bm{p}}=\sqrt{(c\bm{p})^2+(\me c^2)^2}$.

As we see, for the Feynman boundary conditions in the past we have \textit{a priori} an arbitrary combination of only free-particle positron 
states of momentum $-\bm{p}$. On the other hand, in the future we meet again \textit{a priori} an arbitrary superposition of electron states 
of momentum $\bm{p}$, and the well-defined positron state of momentum $-\bm{p}$ and the spin-polarization $\lambda_0$. Arbitrary so far 
constant complex numbers $c^{(+)}_{\bm{p},\lambda}(t_{\mathrm{f}})$ and $c^{(-)}_{\bm{p},\lambda}(t_{\mathrm{i}})$ depend also on $\lambda_0$, 
but in order not to complicate the notation we shall disregard this dependence. Next, as the function $\FeynmanState{\psi}{\bm{p},\lambda_0}{-}(\bm{x},t)$ is supposed to satisfy the Dirac equation for all times, these constants are determined by the transfer matrix,
\begin{equation}
\begin{pmatrix}
c^{(+)}_{\bm{p},+}(t_{\mathrm{f}}) \cr
c^{(+)}_{\bm{p},-}(t_{\mathrm{f}}) \cr
c^{(-)}_{\bm{p},+}(t_{\mathrm{i}}) \cr
c^{(-)}_{\bm{p},+}(t_{\mathrm{i}}) 
\end{pmatrix}
=T_{\bm{p}}(t_{\mathrm{f}},t_{\mathrm{i}})
\begin{pmatrix}
0 \cr
0 \cr
\delta_{+,\lambda_0} \cr
\delta_{-,\lambda_0}
\end{pmatrix},
\label{ewa34}
\end{equation}
and fulfill the normalization condition,
\begin{equation}
1+\sum_{\lambda=\pm}|c^{(+)}_{\bm{p},\lambda}(t_{\mathrm{f}})|^2=\sum_{\lambda=\pm}|c^{(-)}_{\bm{p},\lambda}(t_{\mathrm{i}})|^2 .
\label{ewa35}
\end{equation}
Let us further introduce the real number $N^{(-)}_{\bm{p}}(t_{\mathrm{i}})$ such that
\begin{equation}
[N^{(-)}_{\bm{p}}(t_{\mathrm{i}})]^2=\sum_{\lambda=\pm}|c^{(-)}_{\bm{p},\lambda}(t_{\mathrm{i}})|^2 ,
\label{ewa36}
\end{equation}
and rewrite Eq.~\eqref{ewa35} in the form
\begin{equation}
\sum_{\lambda=\pm} |\mathcal{A}^{(+)}_{\lambda_0}(\bm{p},\lambda)|^2=\sum_{\lambda=\pm} P^{(+)}_{\lambda_0}(\bm{p},\lambda)=P^{(-)}_{\lambda_0}(\bm{p}),
\label{ewa37}
\end{equation}
where 
\begin{equation}
\mathcal{A}^{(+)}_{\lambda_0}(\bm{p},\lambda)=c^{(+)}_{\bm{p},\lambda}(t_{\mathrm{f}}),
\label{ewa38}
\end{equation}
\begin{equation}
P^{(+)}_{\lambda_0}(\bm{p},\lambda)=|\mathcal{A}^{(+)}_{\lambda_0}(\bm{p},\lambda)|^2,
\label{ewa38b}
\end{equation}
and
\begin{equation}
P^{(-)}_{\lambda_0}(\bm{p})=[N^{(-)}_{\bm{p}}(t_{\mathrm{i}})]^2-1.
\label{ewa39}
\end{equation}
According to the Feynman original analysis~\cite{feymann1949qed,feymann1949positrons}, also discussed in the textbook~\cite{bialynicki1975quantumelectro},
$P^{(-)}_{\lambda_0}(\bm{p})$ can be interpreted as the momentum distribution of created positrons of momentum $-\bm{p}$ and spin polarization 
$\lambda_0$. Moreover, $P^{(+)}_{\lambda_0}(\bm{p},\lambda)$ is the momentum distribution of created electrons of momentum $\bm{p}$ and spin polarization 
$\lambda$, provided that the spin polarization of accompanied positrons is $\lambda_0$. As it does depend on the spin state of positrons,
we shall call it the conditional momentum distribution. The same concerns $\mathcal{A}^{(+)}_{\lambda_0}(\bm{p},\lambda)$, which represents the 
corresponding complex conditional amplitudes for the electron momentum distributions. This interpretation follows directly from the scattering 
matrix formalism developed in Ref.~\cite{PhysRevD.110.116025}. Note that since the electric field is homogeneous in space, the momenta of 
created electrons and positrons have opposite space-directions. 


A similar procedure should exists for the case in which electrons are generated in the well-defined states and the conditional amplitudes and 
distributions are determined for accompanying positrons. This goal is achieved if the anti-Feynman boundary conditions are applied.

Following~\cite{bialynicki1975quantumelectro} we say that the state $\AntiFeynmanState{\psi}{\bm{p},\lambda}{+}(\bm{x},t)$ fulfills the anti-Feynman boundary conditions if asymptotically it behaves as
\begin{equation}
\AntiFeynmanState{\psi}{\bm{p},\lambda_0}{+}(\bm{x},t)=
\begin{cases}
\phi^{(+)}_{\bm{p},\lambda_0}(\bm{x},t)+\sum_{\lambda=\pm}\phi^{(-)}_{\bm{p},\lambda}(\bm{x},t),
& t>t_{\mathrm{f}}, \cr 
\sum_{\lambda=\pm}\phi^{(+)}_{\bm{p},\lambda}(\bm{x},t), & t<t_{\mathrm{i}},
\end{cases}
\label{ewa31a}
\end{equation}
where
\begin{align}
\phi^{(+)}_{\bm{p},\lambda_0}(\bm{x},t)&=\ee^{-\ii E_{\bm{p}-\bm{p}_+}(t-t_{\mathrm{f}})}\ee^{\ii\bm{p}\cdot\bm{x}}u^{(+)}_{\bm{p}-\bm{p}_+,\lambda_0} ,
\nonumber \\
\phi^{(-)}_{\bm{p},\lambda}(\bm{x},t)&=\ee^{\ii E_{\bm{p}-\bm{p}_+}(t-t_{\mathrm{f}})}\ee^{\ii\bm{p}\cdot\bm{x}}c^{(-)}_{\bm{p},\lambda}(t_{\mathrm{f}})u^{(-)}_{-(\bm{p}-\bm{p}_+),\lambda} ,
\nonumber \\
\phi^{(+)}_{\bm{p},\lambda}(\bm{x},t)&=\ee^{-\ii E_{\bm{p}-\bm{p}_-}(t-t_{\mathrm{i}})}\ee^{\ii\bm{p}\cdot\bm{x}}c^{(+)}_{\bm{p},\lambda}(t_{\mathrm{i}})u^{(+)}_{\bm{p}-\bm{p}_-,\lambda} .
\label{ewa32a}
\end{align}
Similarly to the Feynman boundary conditions, the unknown coefficients are determined by the inverse of the transfer matrix,
\begin{equation}
\begin{pmatrix}
c^{(+)}_{\bm{p},+}(t_{\mathrm{i}}) \cr
c^{(+)}_{\bm{p},-}(t_{\mathrm{i}}) \cr
c^{(-)}_{\bm{p},+}(t_{\mathrm{f}}) \cr
c^{(-)}_{\bm{p},+}(t_{\mathrm{f}}) 
\end{pmatrix}
=[T_{\bm{p}}(t_{\mathrm{f}},t_{\mathrm{i}})]^{-1}
\begin{pmatrix}
\delta_{+,\lambda_0} \cr
\delta_{-,\lambda_0} \cr
0 \cr
0
\end{pmatrix},
\label{ewa40}
\end{equation}
from which we obtain momentum distributions and corresponding amplitudes for created particles and anti-particles. Namely, if
\begin{equation}
[N^{(+)}_{\bm{p}}(t_{\mathrm{i}})]^2=\sum_{\lambda=\pm}|c^{(+)}_{\bm{p},\lambda}(t_{\mathrm{i}})|^2 ,
\label{ewa41}
\end{equation}
then
\begin{equation}
\sum_{\lambda=\pm} |\mathcal{A}^{(-)}_{\lambda_0}(\bm{p},\lambda)|^2=\sum_{\lambda=\pm} P^{(-)}_{\lambda_0}(\bm{p},\lambda)=P^{(+)}_{\lambda_0}(\bm{p}).
\label{ewa42}
\end{equation}
Here, 
\begin{equation}
P^{(-)}_{\lambda_0}(\bm{p},\lambda)=|\mathcal{A}^{(-)}_{\lambda_0}(\bm{p},\lambda)|^2
\label{ewa43}
\end{equation}
and
\begin{equation}
\mathcal{A}^{(-)}_{\lambda_0}(\bm{p},\lambda)=c^{(-)}_{\bm{p},\lambda}(t_{\mathrm{f}})
\label{ewa43b}
\end{equation}
is the amplitude of created positrons of momentum $-\bm{p}$ and polarization $\lambda$, provided that electrons of momentum $\bm{p}$ and polarization 
$\lambda_0$ are generated with the distribution
\begin{equation}
P^{(+)}_{\lambda_0}(\bm{p})=[N^{(+)}_{\bm{p}}(t_{\mathrm{i}})]^2-1.
\label{ewa44}
\end{equation}

\begin{figure}
\includegraphics[width=7.5cm]{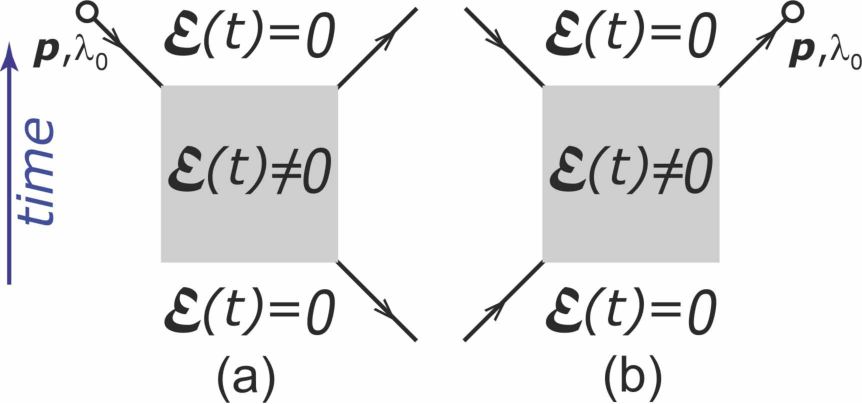}
\caption{Diagrams representing (a) the state $\FeynmanState{\psi}{\bm{p},\lambda_0}{-}(\bm{x},t)$ [cf. Eq.~\eqref{ewa31}] that satisfies the Feynman 
boundary conditions and (b) the corresponding state $\AntiFeynmanState{\psi}{\bm{p},\lambda_0}{+}(\bm{x},t)$ [cf. Eq.~\eqref{ewa31a}] 
for the anti-Feynman boundary conditions (cf. Ref.~\cite{bialynicki1975quantumelectro}, pp. 263 and 265).}
\label{PairFeynman1a}
\end{figure}

The graphical representation of the Feynman and anti-Feynman boundary conditions are shown in Fig.~\ref{PairFeynman1a}, in which the circles
identify those lines that have the well-defined momentum and polarization. Note, however, that for anti-particles the momentum is $-\bm{p}$. 
This representation is based on the corresponding drawings from Ref.~\cite{bialynicki1975quantumelectro}.

\section{Spiral and vortex structures}
\label{sec:Spiralvortex}

Since the seminal research on the spiral structures in photoelectron momentum distributions~\cite{PhysRevLett.115.113004}, different aspects 
of this topic have been investigated. Initially, the existence of spirals was attributed to the creation of quantum vortices. However, the quantum 
vortex has the well-defined mathematical meaning (investigated already by Dirac~\cite{dirac1931quantised}; see, 
also~\cite{PhysRevA.61.032110,Suster:24,PhysRevA.110.023102} and references therein), in contrast to spiral structures, which are only recognized 
in the visual presentation of the corresponding distributions. Based on this fact it was shown that the momentum spirals in multiphoton ionization 
(detachment) are not necessarily related to quantum vortices~\cite{PhysRevA.102.043117}. Complementarily, that quantum vortices are present in distributions 
that do not exhibit the spiral structures~\cite{Larionov2018,PhysRevA.102.043102,PhysRevA.104.033111}. In other words, it was demonstrated that 
spirals and vortices are different physical concepts in quantum theories. Even more, 
by changing the form of the laser pulse it is possible to create or annihilate vortex-anti-vortex pairs without affecting the existence of the spiral 
structures~\cite{majczak2022vorticesphotodetachment}. The point being that the annihilation of the vortex and anti-vortex lines leads to the creation of a nodal 
surface and \textit{vice versa}, which does preserve the separation between the spiral arms.

The Sauter-Schwinger pair creation by an oscillating electric field and the multiphoton ionization have much in common. This has been shown 
by applying the spinorial approach~\cite{bechler2023schwinger,PhysRevD.110.116025,PhysRevD.110.056013,PhysRevD.111.056020}, equivalent 
to the quantum two-level systems. However, the shortcoming of this approach is that it does not fully take into account the spin degrees of freedom 
of both electrons and positrons, and is applicable only for linearly polarized electric fields. We can overcome these problems by using 
the scattering matrix approach~\cite{PhysRevD.110.116025} or, equivalently, the formalism presented in this paper. On the other hand, in the ionization 
of atoms the electron spin does not play an important role and, for this reason, the theoretical analysis mentioned above was carried out using 
the nonrelativistic Schr\"odinger theory. Therefore, the question arises: How does taking into account the spin (helicity) correlations 
of the created electrons and positrons affect the formation of the spiral and vortex structures in the 
momentum distributions? This is the central focus of this section.

\begin{figure}
\includegraphics[width=8.0cm]{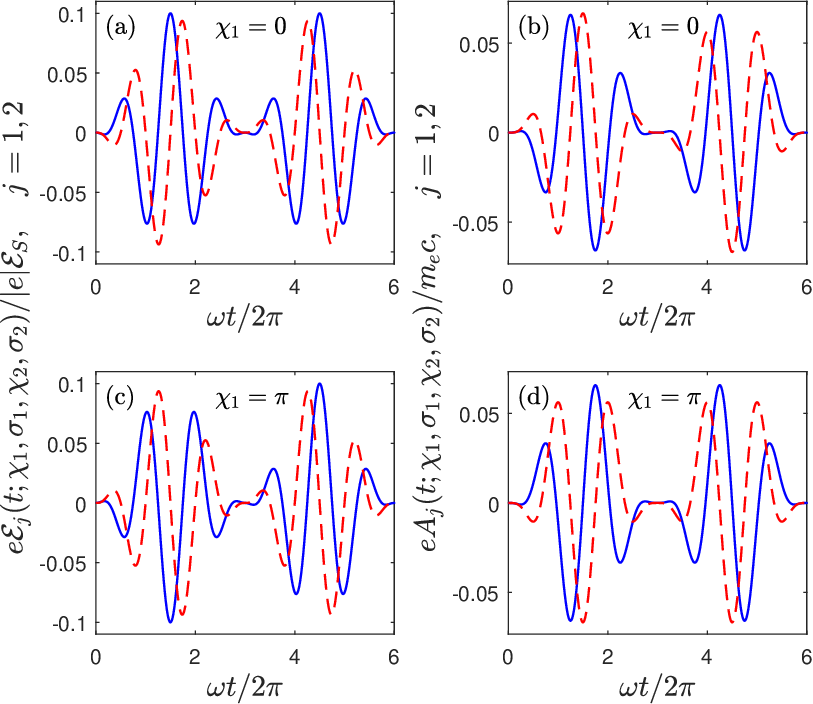}
\caption{Electric field pulses defined by Eq.~\eqref{spin3} [panels (a) and (c)] and the corresponding vector 
potentials~\eqref{spin4} [panels (b) and (d)] for $\mathcal{E}_0=0.1\mathcal{E}_S$, $T_p=2\pi N_\mathrm{osc}/\omega$, 
$N_\mathrm{osc}=3$, $\omega=1.5\me c^2$, $\sigma_1=+1$, $\sigma_2=-1$, $\chi_2=0$, and for two carrier envelope phases of the first pulse: 
$\chi_1=0$ [panels (a) and (b)] and $\chi_1=\pi$ [panels (c) and (d)]. The solid line represents the $x$-coordinate (i.e., $j=1$) of the 
corresponding vector functions, whereas the dashed line relates to the $y$-coordinates (i.e., $j=2$). For these parameters both the electric 
field and the vector potential vanish asymptotically.}
\label{figfields}
\end{figure}

We define an electric field pulse similar to the one used in ionization analysis. In the beginning, let us introduce the envelope function,
\begin{equation}
F(t)=\begin{cases}\sin^2\bigl(\pi\frac{t}{T_p}\bigr), & \textrm{ for } 0\leqslant t \leqslant T_p, \\
            0, & \textrm{ otherwise, }\end{cases}
\label{spin1}
\end{equation}
that acquires nonvanishing values for positive times not longer than $T_p$. This allows us to define the circularly polarized electric field pulse,
\begin{equation}
\bm{\mathcal{E}}_0(t;\chi,\sigma)=\mathcal{E}_0F(t)[\cos(\omega t+\chi)\bm{e}_x+\sigma\sin(\omega t+\chi)\bm{e}_y],
\label{spin2}
\end{equation}
where $\mathcal{E}_0$ is the electric field amplitude, $\omega$ is the frequency of its oscillations, $\chi$ is the carrier envelope phase, 
and $\sigma=\pm 1$ labels the helicity of the electric pulse polarization. Finally, we define the train of two electric pulses,
\begin{equation}
\bm{\mathcal{E}}(t;\chi_1,\sigma_1,\chi_2,\sigma_2)=\bm{\mathcal{E}}_0(t;\chi_1,\sigma_1)+\bm{\mathcal{E}}_0(t-T_p;\chi_2,\sigma_2),
\label{spin3}
\end{equation}
and the corresponding vector potential that vanishes in the far future,
\begin{equation}
\bm{A}(t;\chi_1,\sigma_1,\chi_2,\sigma_2)=\int_t^\infty \dd\tau\,\bm{\mathcal{E}}(\tau;\chi_1,\sigma_1,\chi_2,\sigma_2).
\label{spin4}
\end{equation}
Furthermore, we will assume that each pulse contains $N_\mathrm{osc}$ cycles; hence, $T_p=2\pi N_\mathrm{osc}/\omega$. Two examples of such pulses 
for $N_\mathrm{osc}=3$ and $\mathcal{E}_0=0.1\mathcal{E}_S$ are presented in Fig.~\ref{figfields}, for different values of the carrier-envelope phase 
$\chi_1$. Note that, independently of the field configuration, for $N_\mathrm{osc}\geqslant 2$ both the electric field [Eqs.~\eqref{spin2} 
and~\eqref{spin3}] and the vector potential [Eq.~\eqref{spin4}] vanish in the far past and future.

The theoretical analysis of determining complex probability amplitudes $\mathcal{A}^{(\beta)}_{\lambda_0}(\bm{p},\lambda)$ presented above 
concerned the situation when the spin quantization axis of the created particles is fixed, for instance along the $z$-axis. 
However, by choosing their linear combinations, we can calculate amplitudes for which the electron and positron spin polarizations 
are different and arbitrarily directed in space. In our further analysis we will assume that these axes are determined by the electron and 
positron momenta, which means that we will analyze the amplitudes describing the helicity correlations of created particles. The reason for 
this choice of spin degrees of freedom is that the helicity operator commutes with the free particle Dirac Hamiltonian (see, 
e.g.,~\cite{bialynicki1975quantumelectro,PhysRevD.110.116025} and references therein). This means that the energy and helicity of the electron 
or positron can be measured simultaneously. Therefore, in the further discussion, the quantities $\lambda$ and $\lambda_0$ appearing in the 
amplitudes $\mathcal{A}^{(\beta)}_{\lambda_0}(\bm{p},\lambda)$ always denote the helicities of the particles. Additionally, since the momenta 
of the created electrons and positrons are opposite, without loss of generality only the analysis of electron momentum distributions will be presented, 
which implies the choice of $\beta=+$. This means that $\lambda$ and $\lambda_0$ are the helicities of created electrons and positrons, respectively.

\subsection{Momentum distributions in the $(p_x,p_y)$-plane}
\label{sec:xyplane}

\begin{figure}
\includegraphics[width=8.0cm]{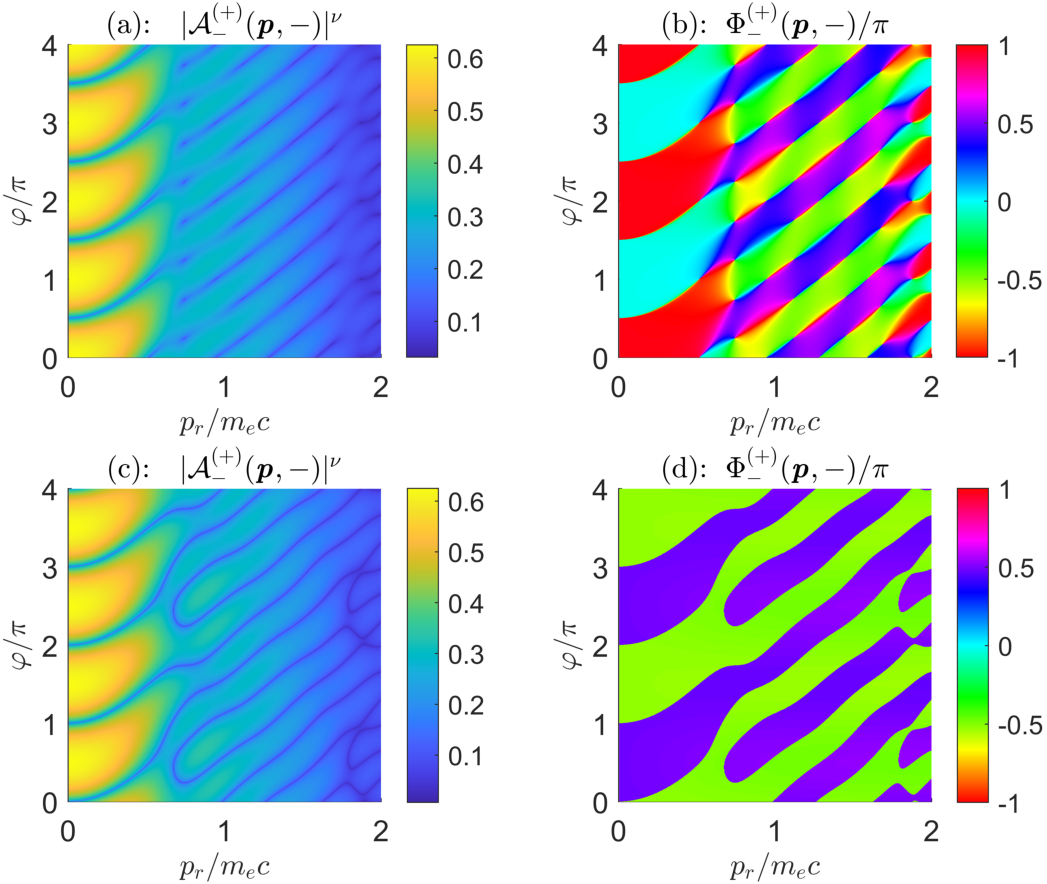}
\caption{Probability amplitudes $\mathcal{A}^{+}_{-}(\bm{p},-)$ for the electric pulses defined in the caption of Fig.~\ref{figfields} 
for $p_z=0$ and for the positron and electron helicities $-1$. Panel (a) presents the modulus of the probability amplitude for $\chi_1=0$, 
raised to the power $\nu=1/4$ for the visual purpose. Panel (b) shows the phase of the amplitude modified by the phase-factor $\exp[\ii\phi(\bm{p})]$ 
with $\phi(\bm{p})$ defined by Eq.~\eqref{spin5}. Panels (c) and (d) represent the same quantities, but for $\chi_1=\pi$.}
\label{figCompareRect4ElectronsK}
\end{figure}

First, we discuss the electron momentum distributions in the $(p_x,p_y)$-plane, i.e. for momenta 
$\bm{p}=p_x\bm{e}_x+p_y\bm{e}_y=p_r(\cos\varphi\bm{e}_x+\sin\varphi\bm{e}_y)$. In Figs.~\ref{figCompareRect4ElectronsK}(a) 
and~\ref{figCompareRect4ElectronsK}(c) we present the moduli of the amplitudes $\mathcal{A}^{(+)}_{-}(\bm{p},-)$ for parameters of the electric 
field pulses defined in the caption to Fig.~\ref{figfields}, for $\chi_1=0$ and $\chi_1=\pi$, respectively. Note that for the visual purpose these 
moduli are raised to the power $\nu=1/4$ and two copies of them are shown in each panel.

A careful analysis of the presented distributions reveals important differences for $\chi_1=0$ and $\pi$. In Fig.~\ref{figCompareRect4ElectronsK}(c) 
(for $\chi_1=\pi$) we observe continuous lines of a practically zero amplitude. For the considered electric field pulse these lines are open, i.e., 
they start and end at the boundaries of the region, even though for longer pulses one can also find closed lines. On the other hand,  
in Fig.~\ref{figCompareRect4ElectronsK}(a) (for $\chi_1=0$) there are lines of small values of the probability amplitude which end abruptly. 
To explain this difference it is necessary to discuss the distribution of the phase of the amplitude (the meaning of the extra phase $\phi(\bm{p})$ 
is discussed below) 
\begin{equation}
\Phi_{\lambda_0}^{(\beta)}(\bm{p},\lambda)=\arg\bigl[\ee^{\ii\phi(\bm{p})}\mathcal{A}_{\lambda_0}^{(\beta)}(\bm{p},\lambda) \bigr],
\label{spin4a}
\end{equation}
presented in Figs.~\ref{figCompareRect4ElectronsK}(b) (for $\chi_1=0$) and~\ref{figCompareRect4ElectronsK}(d) (for $\chi_1=\pi$). The analysis 
of these distributions shows that for $\chi_1=\pi$ the phase $\Phi_{\lambda_0}^{(\beta)}(\bm{p},\lambda)$ at those lines jumps by $\pi$, which proves 
that the amplitude vanishes there. In contrast, for $\chi_1=0$ we observe isolated points for which the amplitude's phase is not defined, i.e., 
in their close vicinity it acquires all possible values from the range $[-\pi,\pi]$. This means that these points are the intersections of vortex 
lines with the $(p_x,p_y)$-plane. For each of these points we can determine the so-called topological charge in such a way that it is a positive 
integer number if during the counterclockwise rotation around this point the phase increases; otherwise, it is negative. The value 
of the topological charge is determined by the number of windings during such a rotation. Analysis of the presented amplitude's phase distribution 
shows that the topological charges of these isolated points are $+1$ (we define this point as a vortex) or $-1$ (anti-vortex). This means that 
the 'lines' of small amplitude values observed in Fig.~\ref{figCompareRect4ElectronsK}(a) are in fact vortex streets, i.e. a sequence of alternating vortices and anti-vortices. Moreover, numerical analysis shows that the change of the phase $\chi_1$ of the electric pulse leads to the change of the position of these isolated vortex points such that for $\chi_1=\pi$ all vortices and anti-vortices are 'annihilated' and the vortex streets turn into continuous lines lying in the $(p_x,p_y)$-plane. Exactly the same scenario is realized in the case of ionization.

\begin{figure}
\includegraphics[width=8.0cm]{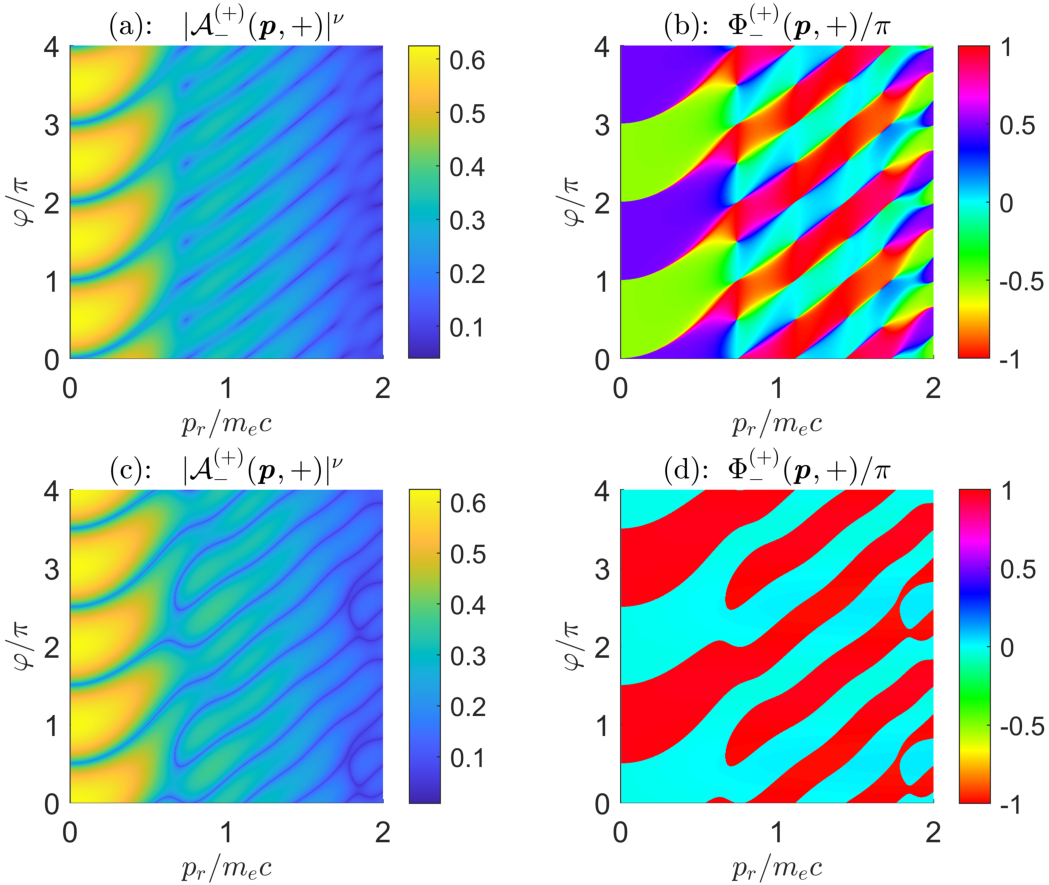}
\caption{The same as in Fig.~\ref{figCompareRect4ElectronsK}, but for the probability amplitude $\mathcal{A}^{+}_{-}(\bm{p},+)$ and for the positron and electron helicities $-1$ and $+1$, respectively.}
\label{figCompareRect2ElectronsK}
\end{figure}

At this point, let us explain the role of the extra phase-factor $\exp[\ii\phi(\bm{p})]$, where we have assumed
\begin{equation}
\phi(\bm{p})=2E_{\bm{p}}T_p=2cT_p\sqrt{\bm{p}^2+(\me c)^2}.
\label{spin5}
\end{equation}
This phase has been used in Eq.~\eqref{spin4a} and in the presentation of the probability amplitude's phase distribution in Fig.~\ref{figCompareRect4ElectronsK}. 
First of all, it does not modify the positions of probability amplitude zeros, but certainly alters its phase pattern. Its purpose is 
to reduce the rapid change of the amplitude's phase and to extract only a slowly changing part of it. In this case, its modulo $2\pi$ part is also a slowly 
changing function of the momentum $\bm{p}$. By doing this all singularities of complex amplitudes, i.e., the jump of the phase by $\pi$ when crossing 
the zero-line and the winding of it around the vortex point are better visualized. Such a procedure concerning the presentation of the amplitude's 
phase distribution has been introduced in Ref.~\cite{PhysRevD.110.116025}.

As mentioned above, the momentum distributions of the created pairs show a significant similarity to the momentum distributions of photoelectrons 
in multiphoton ionization. In the case of long electric field pulses (i.e. when $N_\mathrm{osc} \gg 1$) its Fourier transform is centered around 
the frequency $\omega$. Then the momentum distributions are concentrated around such energies for which $n\omega\approx 2E_{\bm{p}}$, where $n$ 
is a positive integer that can be interpreted as the number of energy quanta absorbed from the electric field during the process. 
The analysis for long pulses shows that in the momentum distributions maxima are observed for
\begin{equation}
p_r=\me c\sqrt{\Bigl(\frac{n\omega}{2\me c^2}\Bigr)^2-1}.
\label{spin6}
\end{equation}
For $n=2$ and $n=3$ we obtain $p_r=1.12\me c$ and $p_r=2.02\me c$, respectively. Indeed, we have checked that for $N_\mathrm{osc}=30$ we observe 
the dominant peak for energy $E_{\bm{p}}=\omega$. Additionally, the number of arms in the spiral distributions grows as $2n$, which is also consistent 
with the results obtained in multiphoton ionization.

Concluding this section, in Fig.~\ref{figCompareRect2ElectronsK} we present the distributions corresponding to the ones shown in 
Fig.~\ref{figCompareRect4ElectronsK}, but for the opposite positron and electron helicities. Although the patterns look very similar, their physical 
meaning is different. This is discussed in the next section.

\begin{figure}
\includegraphics[width=8.0cm]{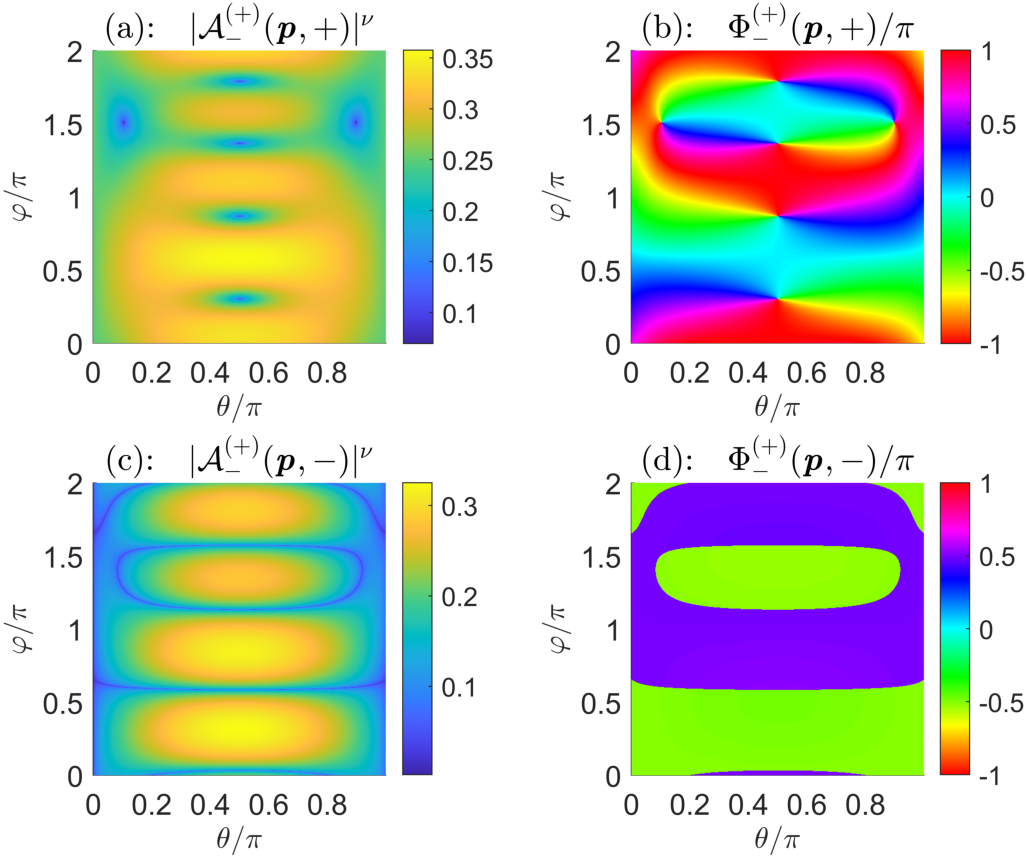}
\caption{The same as in Figs.~\ref{figCompareRect4ElectronsK} and \ref{figCompareRect2ElectronsK} for $\chi_1=\pi$, but on the momentum sphere 
of the radius $p_r=\me c$. These distributions show that the lines of zero probability are either parts of the nodal surfaces 
[Fig.~\ref{figCompareRect4ElectronsK}(c)] or isolated vortex lines [Fig.~\ref{figCompareRect2ElectronsK}(c)].}
\label{figCompareAngles42chi100p1ElectronsK}
\end{figure}

\subsection{Angular distributions for fixed energy}
\label{sec:angles}

In Sec.~\ref{sec:xyplane}, it has been shown that the change of $\chi_1$ can lead to the annihilation of quantum vortices and anti-vortices,
and that the vortex street in the plane of electric field polarization transforms itself into a nodal line. The latter being the intersection 
of the entire nodal surface with the polarization plane. We encounter such a situation in ionization. But this is not the only possible scenario. 
Another possibility is that the vortex street is formed as a result of multiple passages of a wavy vortex line through a selected surface, 
which flattens out while changing the electric field parameters. If it additionally lies on a selected plane, it appears 
as a nodal line. However, it is an isolated line, as it does not belong to the nodal surface. In order to explain which scenario is 
realized for the helicity configurations presented in Figs.~\ref{figCompareRect4ElectronsK} and~\ref{figCompareRect2ElectronsK}, we need to examine 
the distribution of vortices and nodal lines away from the $(p_x,p_y)$-plane. For instance, one can analyze momentum distributions on spheres 
of given momentum values, i.e., for momenta
\begin{equation}
\bm{p}=p_r(\sin\theta\cos\varphi\bm{e}_x+\sin\theta\sin\varphi\bm{e}_y+\cos\theta\bm{e}_z),
\label{spin7}
\end{equation}
with fixed $p_r$ and varied angles $\theta$ and $\varphi$.

\begin{figure}
\includegraphics[width=7.0cm]{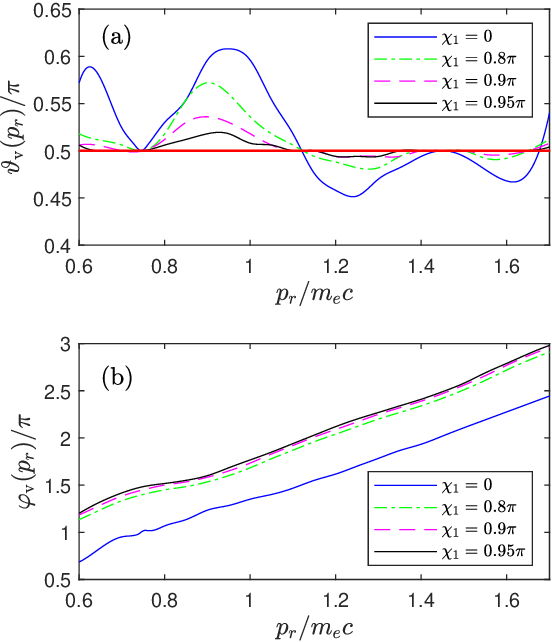}
\caption{Transformation of a vortex line with changing the phase $\chi_1$. The vortex line evolves from the wavy line into the flat one 
lying in the plane $p_z=0$ and presented in Fig.~\ref{figCompareRect2ElectronsK}(c). In upper panel, the straight red line at 
$\vartheta_{\mathrm{v}}(p_r)/\pi=0.5$ corresponds to the case when $\chi_1=\pi$. The remaining values of $\chi_1$ are specified in the legends.}
\label{figvortexlines}
\end{figure}

In Fig.~\ref{figCompareAngles42chi100p1ElectronsK} we present the momentum distributions of the probability amplitude moduli and their phases 
on a sphere of radius $p_r=\me c$ and for $\chi_1=\pi$ for the two spin configurations discussed above. Although in the $(p_x,p_y)$-plane 
(i.e., for $\theta=\pi/2$) these distributions are very similar to each other (cf., Figs.~\ref{figCompareRect4ElectronsK} 
and~\ref{figCompareRect2ElectronsK} for $\chi_1=\pi$), on the sphere they look completely different. For the spin configuration 
$(\lambda_0,\lambda)=(-,-)$ [cf., Figs.~\ref{figCompareRect4ElectronsK}(c) and (d)] there are continuous nodal lines, indicating that these are 
intersections with nodal surfaces. In turn, for the spin configuration $(-,+)$ these are intersections with vortex lines lying in the $\theta=\pi/2$ 
plane. The nodal lines appearing for $\chi_1=\pi$ in Fig.~\ref{figCompareRect2ElectronsK}(c) are therefore vortex lines resulting from the flattening 
out of the wavy vortex lines observed in Fig.~\ref{figCompareRect2ElectronsK}(a) as spiral vortex streets. To see this better, in 
Fig.~\ref{figvortexlines} we present a family of trajectories of a chosen vortex line for several selected values of $\chi_1$ and parametrized 
by the radial momentum $p_r$. The momenta defining this selected vortex line are given by the equation,
\begin{align}\label{spin8}
\bm{p}_{\mathrm{v}}(p_r)=&p_r(\sin[\vartheta_{\mathrm{v}}(p_r)]\cos[\varphi_{\mathrm{v}}(p_r)]\bm{e}_x \\
+&\sin[\vartheta_{\mathrm{v}}(p_r)]\sin[\varphi_{\mathrm{v}}(p_r)]\bm{e}_y+\cos[\vartheta_{\mathrm{v}}(p_r)]\bm{e}_z), \nonumber
\end{align}
with the functions $\vartheta_{\mathrm{v}}(p_r)$ and $\varphi_{\mathrm{v}}(p_r)$ shown in Fig.~\ref{figvortexlines}. 
As one can see, the amplitude of the oscillations of $\vartheta_{\mathrm{v}}(p_r)$ in 
Fig.~\ref{figvortexlines}(a) decreases with increasing $\chi_1$, and for $\chi_1=\pi$ it becomes constant and equal to $\pi/2$. 
Moreover, for $\chi_1$ different from $\pi$, the vortex line is tangent to the surface $\theta=\pi/2$ at the points 
$(p_r,\varphi/\pi)\approx (0.75\me c,1)$ and $(1.45\me c, 2)$, whereas it intersects the surface approximately at the points
$(1.12\me c,1.5)$ and $(1.67\me c, 2.4)$. Such behavior of the vortex line is also visible (even though it is less clear) in 
Fig.~\ref{figCompareRect2ElectronsK}(b).

\begin{figure}
\includegraphics[width=8.0cm]{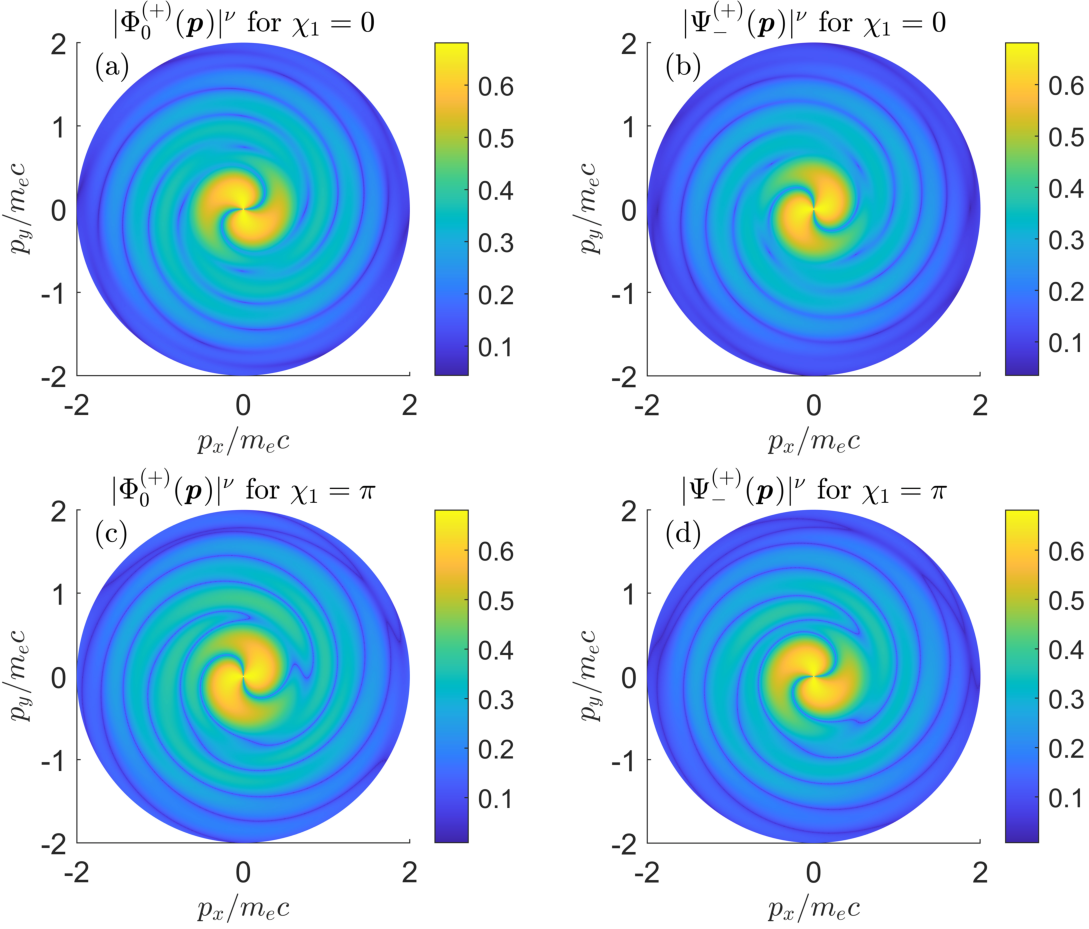}
\caption{Probability amplitudes for the electric pulses defined in the caption to Fig.~\ref{figfields} for $p_z=0$ and for the two nonvanishing 
helicity-entangled probability amplitudes $\Phi^{(+)}_0(\bm{p})$ and $\Psi^{(+)}_-(\bm{p})$ defined in Eq.~\eqref{bell1}. In each panel, the modulus of 
the probability amplitude is raised to the power $\nu=1/4$ for the visual purpose.}
\label{fBellStates000and100ElectronsK}
\end{figure}

\section{Distributions for helicity-entangled states}
\label{sec:Bellstates}

As we have already mentioned, our current approach and the scattering matrix approach~\cite{PhysRevD.110.116025} allow to investigate 
the positron and electron spin-correlation effects in the dynamical Sauter-Schwinger process. This means that, besides the problems discussed 
in Sec.~\ref{sec:Spiralvortex}, we can also study the quantum spin- or helicity-entanglement phenomenon. 
This expands the range of application of the DHW and QKE formalisms. Specifically, because these formalisms provide a partial information about the spin 
effects, as they do not account for the spin degree of freedom of the second particle (in most cases positrons) (see, 
e.g.,~\cite{PhysRevD.99.096017,PhysRevD.100.016013,li2017pairvortex,PhysRevD.107.116010,PhysRevD.111.056020,PhysRevD.110.056013,PhysRevResearch.6.043009,PhysRevD.110.L011901}).

Following the old ideas put forward by Einstein, Podolsky and Rosen~\cite{PhysRev.47.777}, Schr\"odinger~\cite{Schroedinger1935,Schroedinger_1935},
and Bell~\cite{PhysicsPhysiqueFizika.1.195,RevModPhys.38.447} we introduce the helicity-entangled probability amplitudes of pair creation 
(similar to the maximaly entangled Bell states~\cite{Mermin,Bellac} for two spin-$1/2$ systems),
\begin{align}
\Phi^{(\beta)}_0(\bm{p})=&\frac{1}{\sqrt{2}}\bigl[\mathcal{A}^{(\beta)}_-(\bm{p},+)-\mathcal{A}^{(\beta)}_+(\bm{p},-) \bigr], \nonumber \\
\Psi^{(\beta)}_0(\bm{p})=&\frac{1}{\sqrt{2}}\bigl[\mathcal{A}^{(\beta)}_-(\bm{p},+)+\mathcal{A}^{(\beta)}_+(\bm{p},-) \bigr], \nonumber \\
\Psi^{(\beta)}_-(\bm{p})=&\frac{1}{\sqrt{2}}\bigl[\mathcal{A}^{(\beta)}_-(\bm{p},-)-\mathcal{A}^{(\beta)}_+(\bm{p},+) \bigr], \nonumber \\
\Psi^{(\beta)}_+(\bm{p})=&\frac{1}{\sqrt{2}}\bigl[\mathcal{A}^{(\beta)}_-(\bm{p},-)+\mathcal{A}^{(\beta)}_+(\bm{p},+) \bigr].
\label{bell1}
\end{align}
These formulas follow directly from the definition of an arbitrary state of created pairs,
\begin{equation}
|\mathcal{A}^{(\beta)}(\bm{p})\rangle=\sum_{\lambda_0,\lambda=\pm}\mathcal{A}_{\lambda_0}^{(\beta)}(\bm{p},\lambda)|\lambda_0\rangle\otimes|\lambda\rangle,
\label{bell1a}
\end{equation}
and the definition of the Bell states, 
\begin{align}
|\Phi_0\rangle=&\frac{1}{\sqrt{2}}\bigl(|-\rangle\otimes|+\rangle - |+\rangle\otimes|-\rangle\bigr), \nonumber\\
|\Psi_0\rangle=&\frac{1}{\sqrt{2}}\bigl(|-\rangle\otimes|+\rangle + |+\rangle\otimes|-\rangle\bigr), \nonumber\\
|\Psi_-\rangle=&\frac{1}{\sqrt{2}}\bigl(|-\rangle\otimes|-\rangle - |+\rangle\otimes|+\rangle\bigr), \nonumber\\
|\Psi_+\rangle=&\frac{1}{\sqrt{2}}\bigl(|-\rangle\otimes|-\rangle + |+\rangle\otimes|+\rangle\bigr),
\label{bell1b}
\end{align}
where $|\lambda_0\rangle\otimes|\lambda\rangle$ means the helicity product state of created particles. In the following, we shall 
refer to $|\Phi_0\rangle$ as a singlet state, whereas to the remaining states as a triplet of states. Note that
\begin{equation}
\Phi^{(\beta)}_0(\bm{p})=\langle\Phi_0|\mathcal{A}^{(\beta)}(\bm{p})\rangle,
\label{bell1c}
\end{equation}
and similarly for the remaining probability amplitudes.

Fig.~\ref{fBellStates000and100ElectronsK} shows the distributions of the amplitude moduli, $|\Phi^{(+)}_0(\bm{p})|$ and $|\Psi^{(+)}_-(\bm{p})|$, 
for the two carrier-envelope phases of the electric pulse discussed above, $\chi_1=0$ and $\pi$. The remaining distributions defined by Eq.~\eqref{bell1} vanish for $p_z=0$. Similarly 
to the cases discussed in Sec.~\ref{sec:xyplane}, the vortex structures appear explicitly for $\chi_1=0$. However, for $\chi_1=\pi$ they disappear 
for the $\Phi^{(+)}_0(\bm{p})$ distribution (i.e., are transformed into nodal lines, which are part of the nodal surfaces intersecting the $p_z=0$ 
plane), while for the $\Psi^{(+)}_-(\bm{p})$ distribution they evolve with the changing $\chi_1$ into isolated vortex lines in the $p_z=0$ plane. 
The phase distributions of these amplitudes look similar to the ones discussed above and therefore the corresponding figures are not presented.

\begin{figure}
\includegraphics[width=8.0cm]{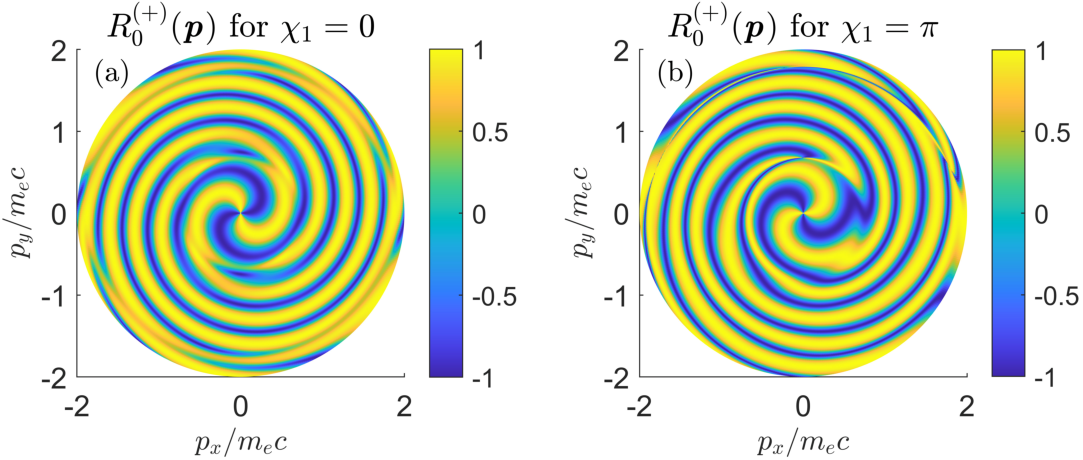}
\caption{The asymmetry momentum distribution defined by Eq.~\eqref{bell2} for two phases $\chi_1$ and for the momentum distributions presented in Fig.~\ref{fBellStates000and100ElectronsK}.}
\label{fBellStates000and100RelElectronsK}
\end{figure}

To investigate the correlations between the generation of pairs in the Bell states $|\Phi_0\rangle$ and $|\Psi_-\rangle$, we introduce 
the asymmetry momentum distribution,
\begin{equation}
R^{(\beta)}_0(\bm{p})=\frac{|\Phi^{(\beta)}_0(\bm{p})|^2-|\Psi^{(\beta)}_-(\bm{p})|^2}{|\Phi^{(\beta)}_0(\bm{p})|^2+|\Psi^{(\beta)}_-(\bm{p})|^2}.
\label{bell2}
\end{equation}
This distribution determines for which momenta $\bm{p}$ the pair is created in one of these entangled helicity states. That is, if for a chosen momentum 
$R^{(\beta)}_0(\bm{p})$ is close to $1$ or $-1$, then the pair appears in the Bell state $|\Phi_0\rangle$ or $|\Psi_-\rangle$, respectively. 
The asymmetry distributions for two chosen electric field phases, $\chi_1=0$ and $\chi_1=\pi$, are presented in Fig.~\ref{fBellStates000and100RelElectronsK}. 
We can see that although they have similar spiral structures, they are however rotated with respect to each other by approximately  
$\pi/2$ (at least for momenta $p_r<\me c/2$). This means that if for a chosen momenta $\bm{p}$ for the phase $\chi_1=0$ the pair is generated 
in the state $|\Phi_0\rangle$, then for the phase $\chi_1=\pi$ it can be created in state $|\Psi_-\rangle$. Thus, the applied electric field can 
be used not only for controlling the generation of selected entangled states, but also can act as a switch between them. This seems to be 
interesting for the rapidly developing area of strong-field quantum simulations, as discussed for instance in~\cite{PhysRevD.109.076004} mainly 
for the Breit-Wheeler process.

\begin{figure}
\includegraphics[width=8.0cm]{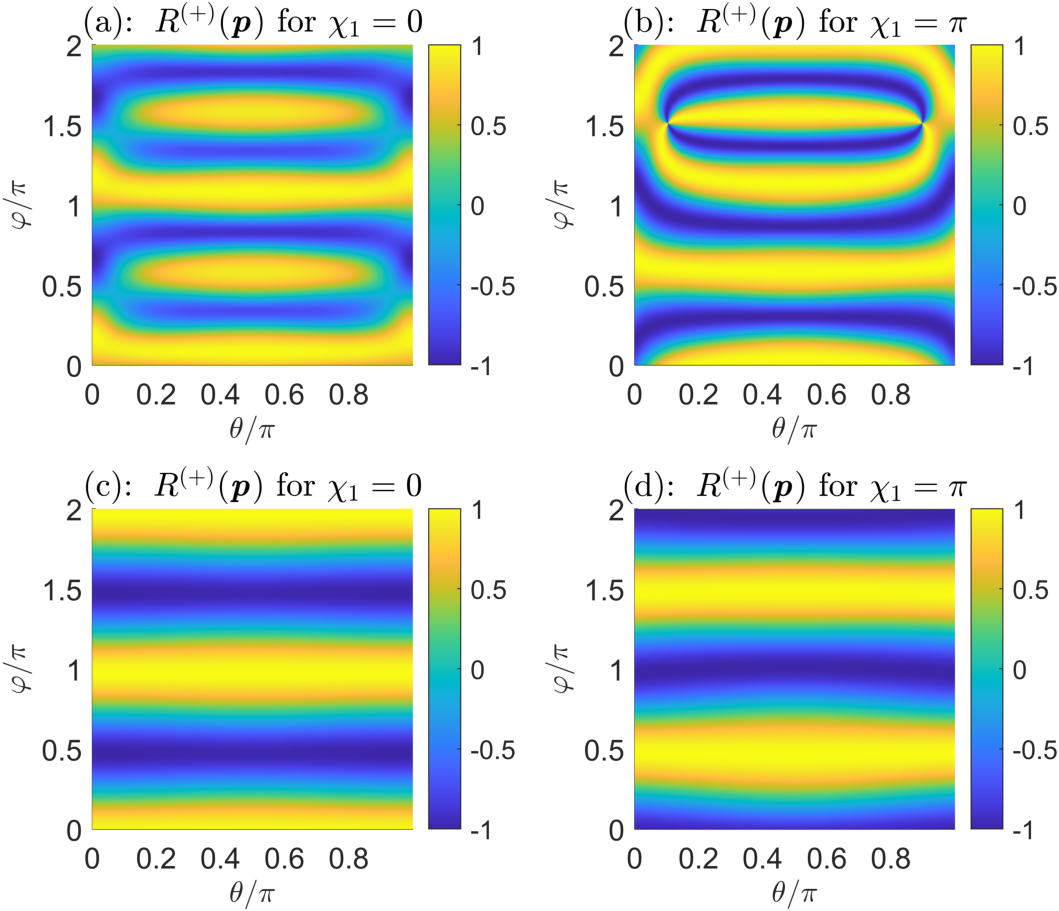}
\caption{The asymmetry momentum distribution defined by \eqref{bell4} for two phases $\chi_1$ and for the momentum distributions calculated on the sphere of momenta \eqref{spin7} for $p_r=\me c$ [panels (a) and (b)], and $p_r=\me c/2$ [panels (c) and (d)].}
\label{fBellStatesAngles000and100Rel2ElectronsK}
\end{figure}

As mentioned above, in the plane $p_z=0$ (and at least for the electric field pulse considered in this paper), pairs are created only in two 
entangled states. In general, however, all states are realized in this process. To take into account such a situation, let us introduce an asymmetry 
distribution that discriminates only between the singlet state $|\Phi_0\rangle$ and three remaining states,
\begin{equation}
R^{(\beta)}(\bm{p})=\frac{|\Phi^{(\beta)}_0(\bm{p})|^2-P^{(\beta)}_{\Psi}(\bm{p})}{|\Phi^{(\beta)}_0(\bm{p})|^2+P^{(\beta)}_{\Psi}(\bm{p})},
\label{bell4}
\end{equation}
where
\begin{equation}
P^{(\beta)}_{\Psi}(\bm{p})=|\Psi^{(\beta)}_0(\bm{p})|^2+|\Psi^{(\beta)}_-(\bm{p})|^2+|\Psi^{(\beta)}_+(\bm{p})|^2.
\label{bell3}
\end{equation}
The corresponding distributions are presented in Fig.~\ref{fBellStatesAngles000and100Rel2ElectronsK}. As we see, for $p_r=\me c/2$ the change of the phase $\chi_1$ from 0 to $\pi$ can serve as the switch between the singlet and triplet entangled states. For larger $p_r$ this goal can not be achieved for the electric field parameters considered here. However, we have checked that by changing the phase $\chi_1$ from 0 to $\pi/2$ the similar switching is available also for $p_r=\me c$.

In summary, we note that the asymmetry distributions do not depend on the way how the initial state is normalized (as mentioned earlier, 
this is an important feature that distinguishes relativistic quantum field theory from condensed matter physics). This means that the analysis presented 
here can be applied to similar problems studied in condensed matter physics, where the equivalent of electron-positron pair production is the excitation 
of an electron to the conduction band and the creation of a hole in the valence band. The unquestionable advantage of such studies is that in this case the electric field 
intensities necessary to observe the above-mentioned effect are experimentally achievable. Therefore, the analysis presented here fits into the general 
trend observed in condensed matter physics, which does not consist in modifying the material medium itself, forcing a change of the band structure in 
the desired manner, but rather in applying an external electromagnetic field (usually laser pulses), which dynamically changes the band structure on 
a very short time scale. The recently popular Floquet engineering is an example of such 
investigations~\cite{uchida2022diabatic,borsch2023lightwave,PhysRevB.108.064308,kim2021ultrafast,popova2024microscopic,bai2022ultrafast,PhysRevB.98.075422,PhysRevA.56.748,fu2025floquet,annurev-conmatphys-031218-013423}. 
Regardless of the above, the study of strong-field QED remains important from the fundamental point of view.

\section{Conclusions}
\label{sec:conclusions}

The Dirac two photon pair annihilation process and the Breit-Wheeler two photon pair creation process inspired Heisenberg and Euler, and later 
Schwinger, to consider the concept of the effective Lagrangian of QED and to investigate the pair creation by a constant electromagnetic fields 
(see, e.g., \cite{ruffini2010pairastro} for historical notes). Direct experimental verification of this phenomenon, however, requires very strong 
fields, which are unattainable experimentally. For this reason, research has recently been conducted on the dynamical Sauter-Schwinger effect with 
the hope that in the near future pulsed electric fields of intensities close to the Schwinger limit will be generated in the laboratory, 
resulting for instance from the concentration of many laser pulses in a small volume and in a short time. Regardless of this, the equivalent
processes to the Sauter-Schwinger effect are already realized in condensed matter physics or photonics. Therefore, studying this phenomenon in 
its 'purest' form, i.e., in relativistic QED, is purposeful, because in this way new approaches or concepts are developed. In this respect, 
the situation encountered in the case of the Sauter-Schwinger effect is 
not unique, as there are many other examples of theoretical research that for many years were not experimentally confirmed (e.g., gravitational 
waves~\cite{RevModPhys.94.025001} or Bose-Einstein condensates~\cite{RevModPhys.71.463}). We also meet a similar situation in the case 
of certain theoretical concepts (e.g., Majorana fermions \cite{10.1093/ptep/ptae065}), which are being revived in other areas of physics.

The aim of this paper was to investigate helicity correlations and entanglement in the electron-positron pair creation by a time-dependent pulsed 
electric field. Also, we have analyzed topological effects appearing in the momentum distributions of created particles. To achieve these goals, 
we have used the approach which is equivalent to the scattering matrix formalism~\cite{PhysRevD.110.116025}, but is more straightforward.
Namely, we have demonstrated that the Sauter-Schwinger 
process can be fully described by the Dirac equation~\cite{bechler2023schwinger} (i.e., without the need for the second quantization formalism) 
with appropriately chosen boundary conditions. We have shown that, for the considered electric field pulses, vortex structures in the momentum
distributions of created particles significantly depend on helicity correlations. This is in contrast to spiral structures, which are only slightly 
modified by changing the helicities of the particles. These conclusions are also valid when the applied electric field changes. Additionally, we have 
investigated the generation of helicity-entangled states and have demonstrated that the electric field can play the role of a switch between 
different entangled states. We suppose that this result can be applied in quantum simulations within strong-field QED for various scenarios of
the electron-positron pair creation, for instance, for the Breit-Wheeler process~\cite{PhysRevD.109.076004,Tang2025} or the trident one~\cite{photonics12040307}. We also assume that these results can be used in the analysis of processes occurring in condensed matter physics.
Finally, we would like to note that for the considered laser field such that $|\bm{\mathcal{E}}(t)|\ll \mathcal{E}_S$, the momentum
distributions presented in this work after summing them up over the helicity (spin) degrees of freedom are nearly identical to the respective
distributions calculated with the DHW or the QKE formalisms.

\section*{Acknowledgements}

We would like to thank E. Saczuk and F. Cajiao V\'elez for discussions.

\bibliography{sf2biblio}

\end{document}